\documentclass[useAMS,usenatbib]{mn2e}
\usepackage{amsmath,fleqn,graphicx,amssymb}
\def\be{\begin{equation}} 
\def\ee{\end{equation}}
\def\Rd{R_{\mathrm d}}
\def\rh{r_{\mathrm h}}

\title{The haloes of merger remnants}

\author[P.~J.~McMillan, E.~Athanassoula \& W.~Dehnen]{
  P.~J.~McMillan$^{1,2,3}$\thanks{E-mail: p.mcmillan1@physics.ox.ac.uk},
  E.~Athanassoula$^{2}$
  and  W.~Dehnen$^{1}$\\
  $^{1}$Department of Physics \& Astronomy, 
  University of Leicester, Leicester, LE1 7RH, UK\\
  $^{2}$LAM, Observatoire Astronomique de Marseille Provence, 2 Place Le Verrier,
  F-13248 Marseille Cedex 4, France\\
  $^{3}$Rudolf Peierls Centre for Theoretical Physics, 1 Keble Road,
  Oxford, OX1 3NP, UK
}

\begin{document}

\maketitle

\begin{abstract}
We perform collisionless $N$-body simulations of 1:1 galaxy 
mergers, using models which include
a galaxy halo, disc and bulge, focusing on the behaviour of the 
halo component. The galaxy models are constructed without recourse 
to a Maxwellian approximation. We investigate the effect of varying the 
galaxies' orientation, their mutual orbit, and the initial velocity
anisotropy or cusp strength of the haloes upon 
the remnant halo density profiles and shape, as well as on the kinematics. 
We observe that the halo density profile (determined as a spherical 
average, an approximation we find appropriate) is exceptionally robust in 
mergers, and that the velocity anisotropy of our remnant haloes
is nearly independent of the orbits or initial anisotropy of the
haloes. The remnants follow the halo anisotropy - local density slope
($\beta-\gamma$) relation suggested by \cite{HansenMoore2006} in the
inner parts of the halo, but $\beta$ is systematically lower than 
this relation predicts in the outer parts. Remnant halo axis ratios are strongly 
dependent on the initial parameters of the haloes and on their orbits.
We also find that the remnant haloes are significantly less spherical 
than those described in studies of simulations which include 
gas cooling.
\end{abstract}


\section{Introduction} \label{sec:intro}

Simulations of structure formation in the $\Lambda$CDM cosmology
paradigm consistently produce triaxial dark matter haloes with density
profiles with $\rho\rightarrow\infty$ as $r\rightarrow0$. This is
commonly referred to as a density ``cusp''. In particular, it is found
that the spherically-averaged density profile in the inner halo
follows a power law, $\rho\propto r^{-\gamma_0}$ with
$\gamma_0\sim1$, while in the outer parts of the halo
$\rho\propto r^{-3}$ \citep[e.g.][]{NavarroFrenkWhite1996,
  MooreEtal1998, PowerEtal2003}. This applies universally over all
mass ranges investigated.

The asphericity of dark matter haloes is also a generic prediction of
$\Lambda$CDM simulations
\citep[e.g.][]{DubinskiCarlberg1991,Allgoodtal2006}, with typical
minor-to-major axes ratios $c/a\sim0.6-0.7$ for a Milky Way sized
galaxy. However, observations of the tidal stream of the Sagittarius
dwarf spheroidal have been used to argue that the halo of the Milky
Way is nearly spherical, with $c/a<0.7$ ruled out to a high degree of
confidence for Galactocentric distances in the range
$16\mathrm{kpc}<r<60\mathrm{kpc}$
\citep{Ibataetal2001,Majewskietal2003}. This is not necessarily a
surprising result, since simulations which take into account the
effect of gas cooling on dark matter haloes find that it results in
haloes that are significantly more spherical than those found in pure
CDM simulations \citep[e.g.][]{KatzGunn1991,Kazanzidisetal2004}.

\cite{HansenMoore2006} found that there is also an apparently universal
relationship between the local density slope and velocity anisotropy
of simulated dark matter haloes.  They argued that this could be understood
through a recognition that the density slope in the tangential
direction was zero (for a spherical halo), and that the shape of a
velocity distribution is directly dependent on the density slope
\citep[which is known for simple structures; e.g.][]{Hansenetal2005}.
The anisotropy of the halo is parameterised by
\begin{equation} \label{eq:beta}
\beta\equiv 1-\frac{\sigma^{2}_{\theta}+\sigma^{2}_{\phi}}{2\sigma^{2}_{r}},
\end{equation}
and the local density slope as
\begin{equation}\label{eq:locg}
\gamma(r)=-\frac{{\mathrm d} \log\rho(r)}{{\mathrm d} \log(r)}.
\end{equation}
\citeauthor{HansenMoore2006} suggested that the anisotropy relates
to the density slope as
\begin{equation}\label{eq:HM}
\beta(\gamma)=1-\xi(1-\gamma/6).
\end{equation}
They used data from collapse, merger and cosmological simulations to
support their hypothesis, and found that the best fit to their data
come from taking the free parameter $\xi\simeq1.15$.
\cite{DehnenMcLaughlin2005} showed that assuming this relationship,
spherical symmetry, and the relationship $\rho/\sigma_r^3\propto
r^{-\alpha}$ where $\alpha$ is some constant
\citep[e.g.][]{TaylorNavarro2001}, defines a single analytical
dynamical equilibrium for a dark matter halo, independent of $\alpha$
 with a density profile
which closely resembles that found in simulations.

Since the work of \cite{ToomreToomre1972} on the tidal origin of
bridges and tails, it has been recognised that elliptical galaxies
could be formed by the merger of two disc galaxies. Various authors
have conducted numerical ($N$-body) simulations of collisionless
mergers, some focusing on the stellar disc component
\citep[e.g.][]{NegroponteWhite1983,Barnes1988,Hernquist1992,NaabBurkert2003},
some on the dark matter halo \citep[e.g.][]{White1978, Villumsen1983,
  BoylanKolchinMa2004, Kazanzidisetal2004, AcevesVelazquez2006,
  KazantzidisZentnerKravtsov2006}.

The studies of White, Villumsen and Boylan-Kolchin \& Ma, 
and most other studies of the halo
component used idealised, spherically symmetric systems to represent
the galaxies. There are relatively few studies of the remnant haloes
of mergers between haloes containing disc galaxies.

\cite{Kazanzidisetal2004} looked at the shapes of dark matter haloes,
primarily considering the effect of gas cooling, but also reporting
simulations of collisionless mergers of galaxies with disc components.
They stated that the shape of the remnant halo depends sensitively on
the relative inclination of the discs.  \cite{AcevesVelazquez2006}
conducted simulations of mergers of galaxies with mass-ratios of 1:1,
1:3 and 1:10, determining that in each case the initial ``cuspy''
density profile was preserved. \cite{KazantzidisZentnerKravtsov2006},
as part of a larger study of dark matter halo mergers, included
simulations of binary (1:1 mass-ratio) mergers of haloes with disc
components. They too found that the presence of a disc did not affect
the conclusion that ``dissipationless mergers result in remnants that
are practically scaled versions of their progenitors''.

All three of these papers create their initial conditions using the
prescription of \cite{Hernquist1993}, which makes the approximation of
a locally Maxwellian velocity distribution, with mean velocity and the
dispersion tensor found from the moment equations.  This method is far
from rigorous as the true equilibrium velocity distribution can be
strongly non-Maxwellian.

\cite{Novaketal2006} looked at the shapes of both the stellar and dark
matter components of the remnants from SPH simulations of
major mergers.  They found that the stellar remnants were generally
oblate, while the haloes were prolate or triaxial. The initial haloes
in their simulations were spherical, and the majority of remnant
haloes were still relatively close to spherical, having $c/a>0.75$,
with no haloes from any of the 58 simulations having $c/a<0.6$.

The most useful observational work on individual dark matter halo
shapes for elliptical galaxies,
\cite{BuoteCanizares1996,BuoteCanizares1998} and \cite{Buoteetal2002},
uses measurements of the flattening of X-ray isophotes to place
constraints on the ellipticity of the dark matter haloes of elliptical
galaxies NGC1332, NGC3923 and NGC720, under the assumption that gas rotation
has a negligible effect. They find that all the haloes
are substantially flattened, with $0.28<c/a<0.65$, and place tight
constraints on the ellipticity of NGC720, which has $c/a =
0.38\pm0.05$ for any of their oblate density models, and $c/a =
0.37\pm0.04$ for any of their prolate models.

No previous simulation studies give a thorough global overview of the properties 
of the haloes of merger remnants, covering their structural and kinematic
aspects. We perform collisionless $N$-body simulations of 1:1 mass 
ratio mergers of
galaxy models consisting of bulge, disc and halo components.  A 
new method for creating a self-consistent equilibrium
model of these components is introduced, which avoids the use of the local Maxwellian
approximation.  We examine the effects of the initial galaxy
orientation, pericentre separation of the mutual parabolic orbit (which we also
refer to as the ``impact parameter''), initial anisotropy of the galaxy
halo, and cusp strength on the density profile, shape, and kinematics
of the remnant halo.

In Sections~\ref{sec:GM}~and~\ref{sec:nummisc} we describe the galaxy
models used in our simulations; in Section~\ref{sec:sims} the various suites of simulations we perform
are detailed; the density profiles, kinematics, and shapes of the remnant haloes 
are examined in Sections~\ref{sec:dens}, \ref{sec:vdisp} and \ref{sec:axes} respectively. 
Finally we discuss our results and draw conclusions.

\section{Galaxy models and simulations} 
\subsection{Initial Conditions} \label{sec:GM}
The problem of building an equilibrium disc-bulge-halo system is a
long standing issue, addressed in many different ways in the
literature \citep[e.g.][]{Hernquist1993,WidrowDubinski2005}.  Our
approach is new, is described in \cite{TheThesis}, and will be
presented in a future paper \citep{McMillanDehneneventually}.  
This approach bears some
relation to that of \cite{Barnes1988}, in that the non-spherically
symmetric component (the disc), is grown adiabatically within the
spheroid. The process has three stages:

\begin{enumerate}
\item Creating an equilibrium initial $N$-body representation of the
  spherically symmetric components of the galaxy (the halo and bulge)
  in the presence of an applied potential field corresponding to the
  monopole (spherical average) of the desired disc potential. The
  distribution function of the halo and bulge is that of
  \cite{Cuddeford1991}. This distribution function has 
\begin{equation} \label{eq:OM}
\beta(r)=\frac{r^2+\beta_0r^2_{\mathrm a}}{r^2+r^2_{\mathrm a}},
\end{equation}
This allows for constant $\beta$ (as \mbox{$r_a\rightarrow\infty$}) or for
  Osipkov-Merritt-model-like anisotropy (with $\beta_0=0$)
  \citep{Osipkov1979,Merritt1985}. 
\item Evolving the $N$-body system whilst growing the non-monopole
  components of the disc potential adiabatically in simulation, 
  allowing halo and bulge
  particles time to relax into the full potential of the disc.
\item Populating the disc component with particles. We use an
  implementation of the disc distribution function $f_{\rm new}$ defined
  in \cite{Dehnen1999:DF}, see Equation~\ref{eq:fnew} of this paper.
\end{enumerate}
 
For the halo we use a spherically-symmetric truncated generalised
Navarro~et~al. (1996, NFW) like profile:
\begin{equation} \label{eq:tNFW}
\rho_{\mathrm h} (r) = \frac{\rho_{\mathrm{c}}}{(r/r_{\mathrm{h}})^{\gamma_0}
(1 + r/r_{\mathrm{h}})^{3-\gamma_0}}   \textrm{sech}(r/r_{\mathrm{t}}),
\end{equation}
where $\rho_{\mathrm{c}}$ is a scale density, $r_{\mathrm{h}}$ is the
halo scale radius, $r_{\mathrm{t}}$ is the halo truncation radius, and
$\gamma_0$ describes the inner slope of the density profile.  As
$r{\rightarrow}0$, the halo density \mbox{$\rho_{\mathrm
    h}\,{\propto}\,r^{-\gamma_0}$}.  We truncate at a radius
significantly larger than the halo scale radius, so in the outer halo
the density goes as \mbox{$\rho_{\mathrm
    h}\,{\propto}\,r^{-3}{\textrm{sech}}(r/r_{\mathrm{t}})$}.  Some
truncation is necessary as the untruncated profile
(\mbox{$r_{\mathrm{t}}\rightarrow\infty$}) is infinitely massive.

\begin{figure}
  \centerline{\resizebox{\hsize}{!}{\includegraphics{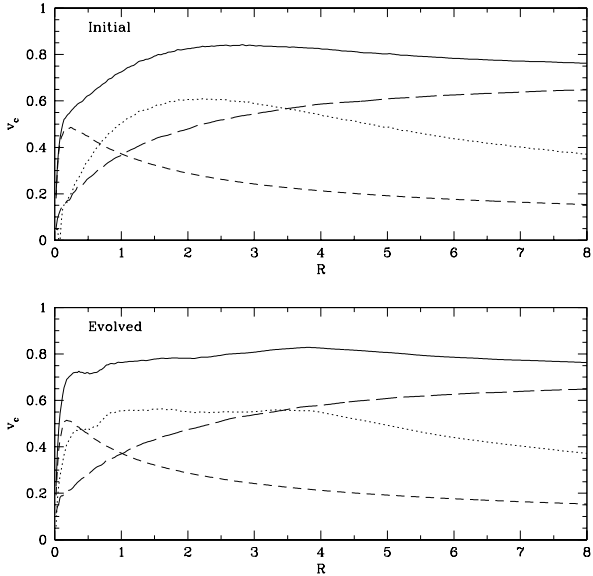}}}
  \caption{Rotation curve in the plane of the disc for 
    our standard $\gamma_0=1$ galaxy model,
    shown for both the initial conditions (\emph{upper}), and after
    being allowed to evolve in isolation for 300 time units (approximately
    the time it takes the galaxies to reach pericentre in our 
    merger simulations -- \emph{lower})
    The solid line is the net rotation curve, also shown are the
    decomposed contributions from the disc (\emph{dotted}), bulge
    (\emph{short-dashed}), and halo (\emph{long-dashed}). The evolution
    seen is dominated by that in the inner $\sim2\Rd$ of the disc component. This is
    caused by a bar instability in the disc which
    redistributes its mass. This, in turn, has some effect on the 
    halo at its innermost radii (see Section~\ref{sec:sims}).
   \label{fig:rotcurve}
 }
\end{figure}

The disc is defined as having a surface density profile which is
exponential in the cylindrical radius $R$, and a
vertical ($z$-component) structure modelled by isothermal sheets
\begin{equation} \label{eq:discrho}
\rho_{\mathrm d}(R,z) = \frac{M_{\mathrm d}}{4\pi R_{\mathrm d}^2 z_{\mathrm d}}
\textrm{exp}\left(-\frac{R}{R_{\mathrm d}}\right)
{\textrm {sech}}^2\left(\frac{z}{z_{\mathrm d}}\right),
\end{equation}
where $M_{\mathrm d}$ is the total disc mass, $\Rd$ is the disc scale
radius, and $z_{\mathrm d}$ the scale height.

To find the distribution function of the disc, we assume that it can
be decomposed into its components in the plane of the disc, and
perpendicular to it, i.e.  \mbox{$f_{\mathrm d}\cong f_{\mathrm d}(E_p,L_z,E_z)$} 
where the ``vertical(z-component) energy'' 
\mbox{$E_z \equiv \frac{1}{2} v_z^2 + \Phi(R,z)  - \Phi(R,0)$}, 
and the ``planar energy'' \mbox{$E_p =E-E_z=\frac{1}{2} (v_R^2+v_\theta^2)+
\Phi(R,0)$}.

If we define $R_{Ep}$ as the radius of a circular orbit in the disc
midplane with planar energy $E_p$, we can write the planer component
of the disc distribution function as
\begin{multline}\label{eq:fnew}
f_{d,p}(E_p,L_z) =  \\
\frac{\Omega(R_{Ep})\,\Sigma'(R_{Ep})}{\pi\,\kappa(R_{Ep})
\sigma_R'^{2}(R_{Ep})}\,
\exp\!\left[\frac{\Omega(R_{Ep})[L_z{-}L_{z,c}(E_p)]}{\sigma_R'^{2}(R_{Ep})}\right],
\end{multline}
where $L_{z,c}(E)$ is the angular momentum of a circular orbit in the
disc midplane with planar energy $E_p$, $\Omega$ is the circular frequency
of that orbit, and $\kappa$ is the epicycle frequency. $\Sigma'(R)$
and $\sigma_R^{\prime2}(R)$ are defined such that the true surface
density $\Sigma(R)$ and radial velocity dispersion $\sigma_R^2(R)$ of
the $N$-body representation are those desired, to within an
appropriate degree of accuracy.

We model the bulge as non-rotating, initially spherically symmetric
and with a \cite{Hernquist1990} density profile.
\begin{equation} \label{eq:bulgerho}
   \rho_{\mathrm b}(r) = \frac{M_{\mathrm b}}{2\pi}\frac{r_{\mathrm b}}{r(r_{\mathrm b}+r)^3},
\end{equation}
where $M_{\mathrm b}$ is the mass of the bulge, and $r_{\mathrm b}$ is
its scale radius.

We performed a suite of simulations in an effort to investigate the
effect of the properties of the progenitors and of the encounter
on the halo of the merger of two equal-mass galaxies. The
galaxy models were designed to loosely resemble the Milky Way as
described by \cite*{KlypinZhaoSomerville2002}.

\subsection{Numerical Miscellanea}\label{sec:nummisc}
In all our models we choose units such that the disc scale length, \mbox{$R_{\mathrm d} =
1$}, disc mass \mbox{$M_{\mathrm d} = 1$}, and the constant of gravity
$G = 1$. The disc scale height was chosen to be \mbox{$z_{\mathrm  d} = 0.1$},
the velocity dispersion of the disc was 
defined such that the \cite{Toomre1964} stability parameter $Q=1.2$ at
all radii.
The bulge mass \mbox{$M_{\mathrm b} = 0.2$}, and bulge scale
length \mbox{$r_{\mathrm b} = 0.2$}. Scaling these values to the Milky Way, taking
\mbox{$R_d=3.5$kpc}, \mbox{$M_d + M_b=5 \times 10^{10} $M$_\odot$} gives a
time unit~\mbox{$\simeq 14$Myr}.  We chose to have the
scale radius of the halo \mbox{$r_{\mathrm h} = 6$}, the truncation radius
\mbox{$r_{\mathrm t} = 60$}, and the halo mass \mbox{$M_{\mathrm h} = 24$}. 
In simulations with $\gamma_0=1.0$, 79\% of
this total mass is within the truncation radius. The rotation curve
for this model is shown in Figure~\ref{fig:rotcurve}.

The stellar components were populated with
150\,000 equal mass particles (i.e.~125\,000 in the disc, 25\,000 in
the bulge) with a smoothing length $\epsilon_{\mathrm stellar} = 0.02$. The halo component
was populated with 750\,000 equal mass particles with a smoothing length
$\epsilon_{\mathrm halo} = 0.04$. That corresponds to each halo particle being
4 times more massive than a stellar particle. A small number of simulations
with 4 times more particles were performed for comparison. These demonstrated 
that the numerical resolution used was sufficient.

The $N$-body simulations were performed using the publicly available
$N$-body code \textsf{gyrfalcON}, which is based on Dehnen's
(\citeyear{Dehnen2000, Dehnen2002}) force solver
\textsf{falcON}, a tree code with mutual cell-cell interactions and
complexity $\mathcal{O}(N)$.  The equations of motion were integrated
using the familiar leap-frog integrator with minimum time step
$2^{-7}$ and a block-step scheme allowing steps up to eight times
larger. Individual particle time steps were adjusted in an (almost)
time-symmetric fashion such that on average
\begin{equation}
  \tau_i = \min\left\{\frac{0.01}{|\bmath{a}_i|},\;
    \frac{0.05}{|\Phi_i|}\right\},
\end{equation}
with $\Phi_i$ and $\bmath{a}_i$ the gravitational potential and
acceleration of the $i$th body. With these parameters, energy was
conserved to within 0.1\% over the full time span (1000 time units) in
a typical simulation, approximately corresponding to a Hubble time.
The time span was chosen such that there is sufficient time for the 
remnant to reach a dynamical equilibrium.  

In all simulations the equal mass galaxies were placed on a mutual
orbit in the x-y plane. In the vast majority of cases these orbits
were parabolic and corresponded to one that would have a pericentre
separation of 8 $R_{\mathrm d}$ if the galaxies were point masses; 
we refer to this distance
as the ``impact parameter'' $d$ of the merger. In some cases
the galaxies were placed on orbits with smaller pericentre
separations, or on a radial orbit with zero net energy (i.e. a
parabolic orbit in the limit where impact parameter $d
\rightarrow 0$).  The galaxy centres were initially separated by $200
\Rd$.

\subsection{Suites of simulations} \label{sec:sims}

We perform four suites of simulations of equal-mass mergers, to
examine the effects of varying different parameters on the mergers.
The suites vary in

\begin{enumerate}
\item Orientation of the disc galaxy.
\item Pericentre separation of the galaxies' mutual orbit.
\item Initial velocity anisotropy of the halo component.
\item Cusp strength of the haloes ($\gamma_0$ in Equation~\ref{eq:tNFW}).
\end{enumerate}

In addition, we ran control simulations, which were stand-alone simulations of
the same galaxy models, run for the same length of time, with no
merger. The galaxy disc is bar-unstable, which in these simulations
causes a steepening in the density profile of the inner parts of the
disc, which also causes a steepening in the halo profile in the inner
$\sim0.2\rh$. This steepening is commonly observed in simulations of 
barred galaxy evolution \citep[reviewed e.g. by][]{Athanassoula2004}.
The effect of this evolution on the rotation curve of the galaxy can
be seen in Figure~\ref{fig:rotcurve}. The anisotropy of the halo 
velocity distribution is essentially unchanged by the stand-alone
evolution.

\subsubsection{Suite 1: Orientation} \label{sec:sims:orient}

\begin{figure}
  \centerline{\resizebox{\hsize}{!}{\includegraphics{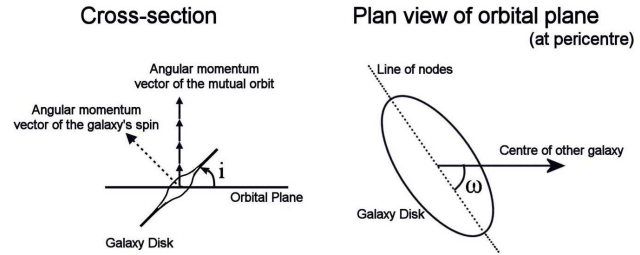}}}
  \caption{Diagram showing the definitions of inclination $i$, and
        argument of pericentre $\omega$ for the mergers. By convention,
        \mbox{$-180^\circ<i\leq180^\circ$}, \mbox{$-90^\circ<\omega<90^\circ$}. The 
        ``line of nodes'' is the intersection of the plane of the galaxy 
        disc with the orbital plane. The cross section view (\emph{left})
        is that from a point in both the plane of the orbit (shown as the
        horizontal), and the plane of the galaxy disc (shown at an angle
        $i$ to the horizontal). The plan view is from above the orbital plane, 
        looking down.
   \label{fig:argperi}
  }
\end{figure}

Our first suite of simulations consisted of 16 mergers which vary only
in the orientation of the discs with respect to one another, and to
the angular momentum vector of their mutual orbit. Following the
example of \cite{Barnes1988}, we define four different orientations for
the spin vector of each galaxy disc, corresponding to pointing towards
the four vertices of a regular tetrahedron.  The orientations of disc
galaxies in mergers are usually described in terms of the disc
inclination relative to the orbital plane $i$, and the argument of
pericentre $\omega$ 
\citep[See also Figure~\ref{fig:argperi} or][]{ToomreToomre1972}. 
The inclinations
($i_{\mathrm 1}\,\&\,i_{\mathrm 2}$), and arguments of pericentre
($\omega_{\mathrm 1}\,\&\,\omega_{\mathrm 2}$) for the suite of mergers
are shown in Table~1.

These are the same initial orientations that were used in the collisionless merger simulations
of \cite{NaabBurkert2003} and the SPH merger simulations of 
\cite*{Naabetal2006}. These studies detailed the significant effects that varying the 
initial disk orientations has upon the properties 
of the \emph{luminous} components of the merger remnants.

In all cases the initial halo has an inner density slope $\gamma_0 = 1.0$
(Equation~\ref{eq:tNFW}), and an isotropic ($\beta=0$) velocity distribution.
The galaxies were put on a parabolic orbit with a pericentre
separation of $8\Rd$.

\begin{tabular}{lcc|cc} 
\multicolumn{5}{c}{}\\
\multicolumn{5}{c}{TABLE 1} \\ \hline
& \multicolumn{2}{c|}{Disc 1}
& \multicolumn{2}{c}{Disc 2} \\ \hline \hline
$\#$ &$i_1$ & $\omega_1$ & $i_2$ & $\omega_2$  \\ \hline
1    & 0    & 0          & 180   & 0   \\
2    & 0    & 0          & 71    & 30  \\
3    & 0    & 0          & 71    & -30 \\
4    & 0    & 0          & 71    & 90  \\
5    & -109 & -60        & 180   & 0   \\
6    & -109 & -60        & 71    & 30  \\
7    & -109 & -60        & 71    & -30 \\
8    & -109 & -60        & 71    & 90  \\
9    & -109 &   0        & 180   & 0   \\
10   & -109 &   0        & 71    & 30  \\
11   & -109 &   0        & 71    & -30 \\
12   & -109 &   0        & 71    & 90  \\
13   & -109 &  60        & 180   & 0   \\
14   & -109 &  60        & 71    & 30  \\
15   & -109 &  60        & 71    & -30 \\
16   & -109 &  60        & 71    & 90  \\ \hline
\end{tabular}

\subsubsection{Suite 2: Impact parameter} \label{sec:sims:impact}
We performed two sets of simulations with galaxy models identical to
those used in Suite 1, varying the separation at pericentre (impact
parameter) $d$ between $8R_{\mathrm d}$ and $0$ (i.e. radial impact) 
in increments of 2$R_{\mathrm d}$. The two sets of simulations differ in the galaxy
orientations, which correspond to those of simulations 4 \& 7 in suite
1 (Table~1).

\subsubsection{Suite 3: Halo anisotropy} \label{sec:sims:ani}
Our third suite of simulations consisted of mergers of galaxies with
varying initial halo velocity dispersions and were performed to examine
the effect that has on remnant properties. The haloes had identical
density profiles to those in suites 1 and 2, but rather than an
isotropic velocity distribution, they were initialised with constant
radial anisotropy (\mbox{$\beta=0.35$} throughout the halo), constant
tangential anisotropy (\mbox{$\beta=-0.4$} throughout the halo), or
Osipkov-Merritt-model-like anisotropy (Equation~\ref{eq:OM}) with
\mbox{$r_{\mathrm a} = 24 = 4r_{\mathrm h}$}. Simulations were performed 
with galaxies orientated as per orientation 7 (Table~1),
and with impact parameters of 8, 3, and $0\Rd$.

\subsubsection{Suite 4: Cusp strength} \label{sec:sims:cusp}
We ran a series of simulations with varying halo cusp strengths.
The inner slope of the density distribution ($\gamma_0$ in
Equation~\ref{eq:tNFW}) varied between \mbox{$\gamma_0=0.1$} and \mbox{$\gamma_0=1.6$}. In all
cases we retained the parameters \mbox{$r_{\mathrm h}=6$}, 
\mbox{$r_{\mathrm t}=60$} and \mbox{$M_{\mathrm h} = 24$}. The galaxy 
discs had
the same orientations as simulation 4 in Table~1. The haloes
all had isotropic velocity distributions, and the mergers trajectories
all had impact parameter $d=8\Rd$.

\section{Results}
\subsection{Halo density profiles} \label{sec:dens}

The haloes of the merger remnants are not spherically symmetrical (see
Section~\ref{sec:axes}), however insight, and a comparison with previous 
work, can be found from looking at
spherically averaged density profiles, and local density slopes
(Equation~\ref{eq:locg}). 

\begin{figure}
  \centerline{\resizebox{\hsize}{!}{\includegraphics{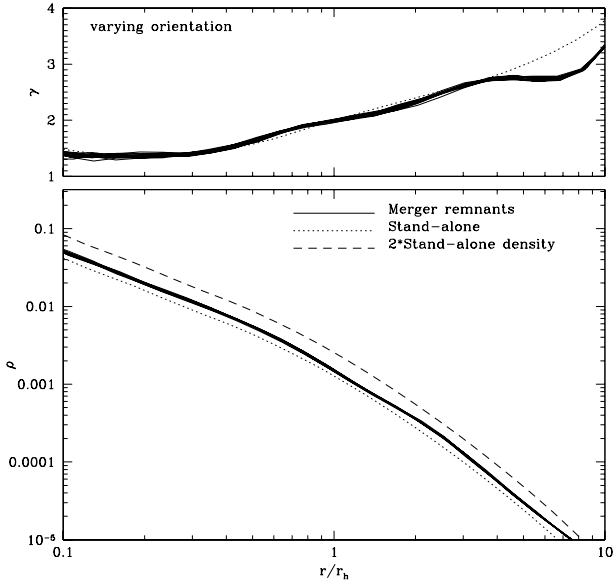}}}
  \caption{Suite 1: Spherically averaged density profile (\emph{lower})
    and density slope profile ($\gamma$, see Equation~\ref{eq:locg};
    \emph{upper}) for remnant haloes from the 16 simulations in suite
    1 (\emph{solid lines}).  Also shown, for comparison, is the
    density profile and slope of the halo of the equivalent
    stand-alone simulation (\emph{dotted}), and the same stand-alone
    density profile with the density doubled - i.e. what one would
    expect if the galaxies were simply stacked on top of one another
    (\emph{dashed}).
   \label{fig:orientpro}
 }
\end{figure}

Figure \ref{fig:orientpro} shows the density profiles and slopes for
the 16 merger remnants from suite 1 (Section~\ref{sec:sims:orient}),
with the equivalent for the stand-alone simulations. The disc
orientations have no effect on the density profile, even in the
regions where they dominate the galaxies' density.  The shape of the remnant
density profile is very similar to that of the initial haloes, with an
almost identical scale length. There are, however differences in the 
outer parts: $\sim6\%$ of the halo mass becomes
unbound during the merger, and the mass within $10\rh$ is $\sim50\%$ of
the total mass of the original haloes (as compared to $79\%$ in each
halo initially). Hence the density within the inner $\sim10\rh$ of the
merged haloes is closer to that of one of the initial haloes than to
double that (as it would be if the merger was equivalent to simply
``freezing'' the haloes, and placing them on top of one another).

\begin{figure}
  \centerline{\resizebox{\hsize}{!}{\includegraphics{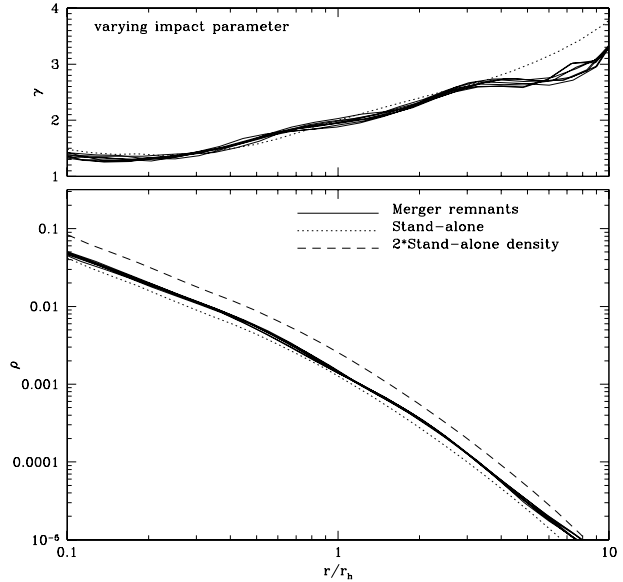}}}
  \caption{Suite 2: Spherically averaged density profile and density 
    slope profile for remnant haloes, as in
    Figure~\ref{fig:orientpro}, for simulations with impact parameters
    \mbox{$d=8,6,4,2,0\Rd$} (\emph{solid lines}).
   \label{fig:imppro}
 }
\end{figure}

\begin{figure*}
  \centerline{\hfil
    \resizebox{82mm}{!}{\includegraphics{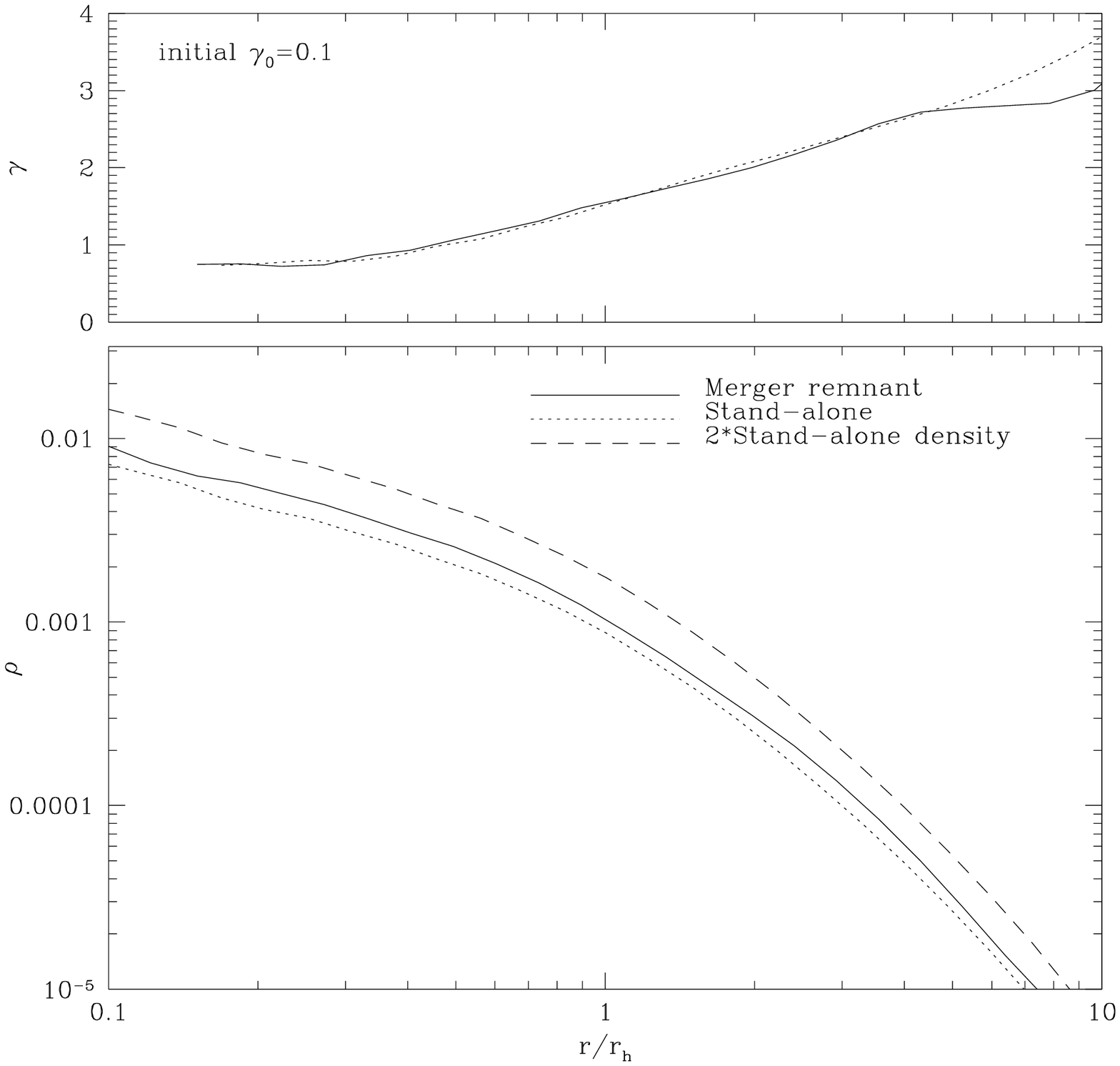}}\hspace{4mm}
    \resizebox{82mm}{!}{\includegraphics{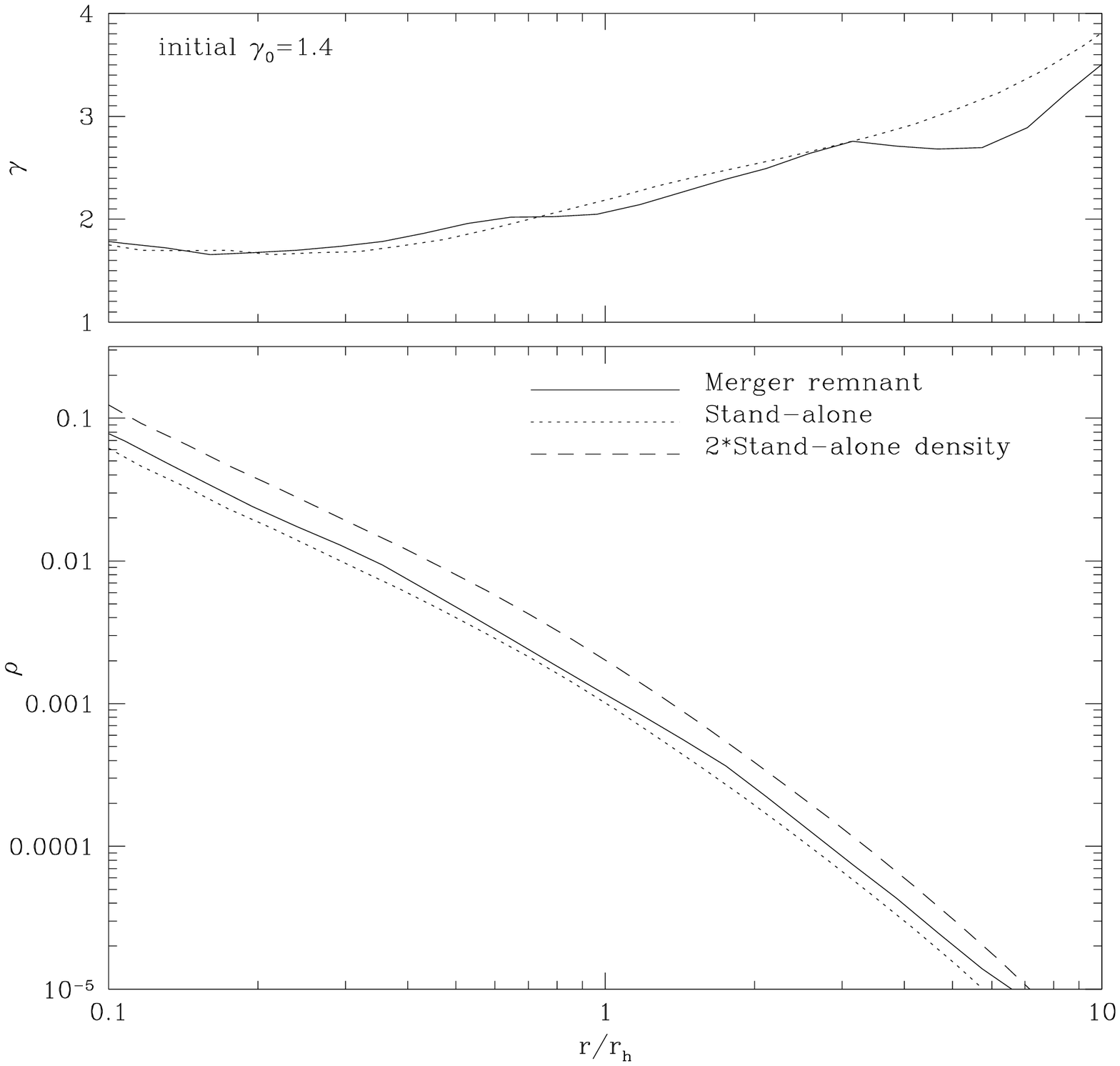}} }
  \caption{Suite 4: Spherically averaged density profile and density 
    slope profile for stand-alone and remnant haloes, as in
    Figure~\ref{fig:orientpro}. The initial haloes had
    \mbox{$\gamma_0=0.1$} (\emph{left}) or \mbox{$\gamma_0=1.4$}
    (\emph{right}) density cusps.
   \label{fig:gams}
 }
\end{figure*}

The simulations of suite 2 show that varying the impact parameter of
mergers also has little effect on the shape of the density profile
(Figure~\ref{fig:imppro}). Results for suite 4 showed the same trends
for all cusp strengths. In the interests of brevity, only the results
for \mbox{$\gamma_0=0.1$} and \mbox{$\gamma_0=1.4$} are shown
(Figure~\ref{fig:gams}).

\begin{figure}
  \centerline{\resizebox{\hsize}{!}{\includegraphics{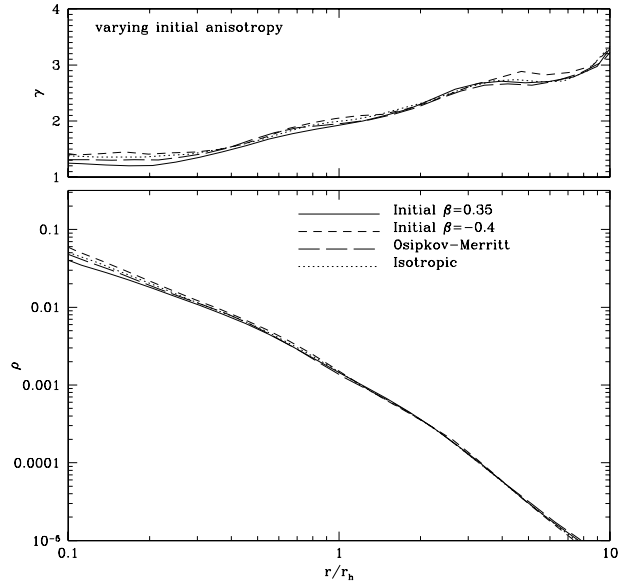}}}
  \caption{Suite 3: Spherically averaged density profile and density 
    slope profile for remnant haloes from \mbox{$d=8\Rd$} mergers in
    which the initial haloes had constant radial anisotropy
    \mbox{($\beta=0.35$:} \emph{solid line}); constant tangential
    anisotropy \mbox{($\beta=-0.4$:} \emph{short-dashed});
    Osipkov-Merritt anisotropy \mbox{($r_{a}=24$:} \emph{long-dashed})
    Also shown are the density profile and slope for the equivalent
    mergers with initially isotropic haloes, as in all other
    simulations \emph{dotted}).
        \label{fig:anipro}
      }
\end{figure}

Results from suite 3 do show a slight trend in remnant cusp strength
with initial halo anisotropy. The trend is observable for all orbital
separations simulated, but only results from simulations with orbital
separation \mbox{$d=8\Rd$} are shown in Figure~\ref{fig:anipro}, for
clarity. Haloes which have initially constant tangential anisotropy
have sharper remnant cusps than initially isotropic haloes, while
haloes with initial constant radial anisotropy have weaker remnant
cusps. Giving the haloes Osipkov-Merritt anisotropy does not alter the
cusp strength.  Compared to the isotropic case, cusp strength
--~measured in terms of $\gamma(r)$ at small radii~-- is
\mbox{$\sim0.1$} greater for the \mbox{$\beta=0.35$} case than for the
isotropic case, and is \mbox{$\sim0.1$} less in the
\mbox{$\beta=-0.4$} case. This trend is not seen in the stand-alone
profiles, which are nearly identical in all three cases.

\subsection{Halo velocity distributions} \label{sec:vdisp}

\begin{figure}
  \centerline{\resizebox{\hsize}{!}{\includegraphics{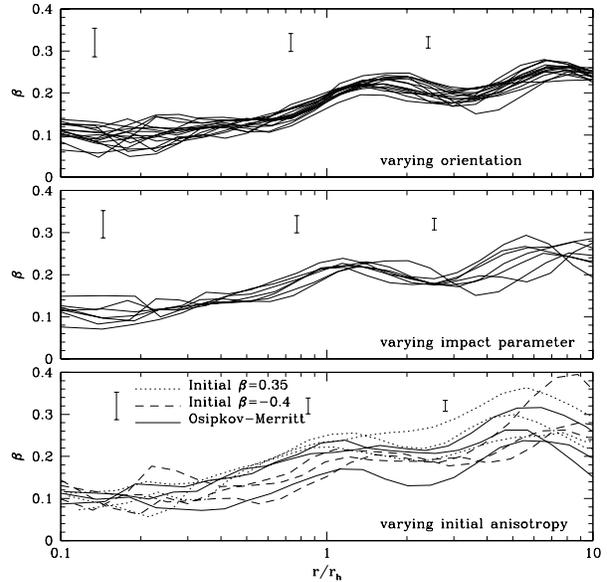}}}
  \caption{$\beta$ as a function of radius for the simulations of suites 1--3 
        (\emph{top to bottom}). Error bars shown are typical of all simulations
        in respective plot, found using Equation~\ref{eq:betaerrs}. 
        In the bottom plot remnants of mergers 
        with haloes with different initial anisotropies are represented by 
        different line types. \emph{Dotted} lines plot $\beta$ for remnants
        of initial halo $\beta=0.35$ mergers,  \emph{dashed} lines for 
        initial halo $\beta=-0.4$, \emph{solid} lines for the Osipkov-Merritt
        haloes.
   \label{fig:oiabeta}
  }
\end{figure}

\begin{figure}
  \centerline{\resizebox{\hsize}{!}{\includegraphics{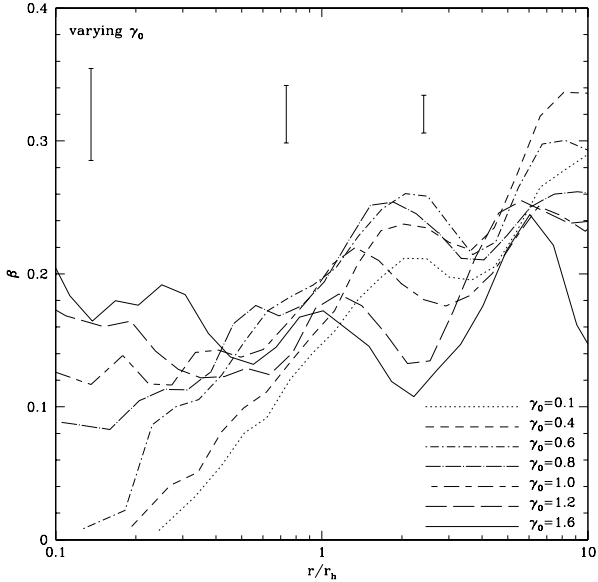}}}
  \caption{$\beta$ as a function of radius for the simulations of suite 4. 
        Error bars shown are typical.
        Simulation results shown are for haloes with density cusps of strengths
        as indicated. 
   \label{fig:cuspbeta}
  }
\end{figure}

We investigate the velocity distribution of the haloes by looking at
$\beta$ as a function of radius, and as a function of the local
density slope $\gamma(r)$. $\beta$ is determined for particles in spherical shells,
with the median particle radius for the shell being the value plotted in 
Figures~\ref{fig:oiabeta}~\&~\ref{fig:cuspbeta}.

Error estimates, as shown in Figures~\ref{fig:oiabeta}~and~\ref{fig:cuspbeta} were
found for the spherical shell of particles
through the standard propagation of errors formula, such that
the error $\Delta\beta$ is given by
\begin{multline}\label{eq:betaerrs}
(\Delta\beta)^2=\\
\left(\frac{\partial\beta}{\partial\sigma_\theta^2}\right)^2
(\Delta\sigma_\theta^2)^2 + \left(\frac{\partial\beta}{\partial\sigma_\phi^2}\right)^2
(\Delta\sigma_\phi^2)^2 + \left(\frac{\partial\beta}{\partial\sigma_r^2}\right)^2
(\Delta\sigma_r^2)^2,
\end{multline}

The remarkable aspect of the observed $\beta$ profiles of the remnant
haloes is its near independence from the parameters varied in suites
1--3 (Figure~\ref{fig:oiabeta}, top to bottom).  In all cases 
examined \mbox{$\beta\approx0.1$} at small $r$,
and increases to \mbox{$\beta\sim0.2-0.3$} at \mbox{$r\sim10r_{\mathrm h}$}.  This
shows that while the remnant halo has a
clear ``memory'' of the density profiles of the initial haloes, it has
little memory of the velocity distribution of the initial haloes.
There is some difference observable between the $\beta=0.35$ and 
$\beta=-0.4$ cases, perhaps most noticeable in the range 
$0.5\rh<r<\rh$, but it is vastly smaller than the difference in the 
initial halo conditions. 

One would also naively expect the impact parameter of the merger to
have a significant effect, but it does not show any greater effect on
the $\beta$ profile than the orientation of the galaxies does.  The
independence from disc orientation is unsurprising at large radii, but
also continues to small radii, where one might have expected the disc
orientation to play a role.

The $\beta$ profiles from the simulations of suite 4 are shown in
Figure~\ref{fig:cuspbeta}.  As \mbox{$r\rightarrow0$}, the value of $\beta$
tends to $0$ for the less strongly cusped initial conditions
\mbox{($\gamma_0=0.1-0.6$)}. As the cusp strength increases, the radial
anisotropy of the inner parts of the halo increases. At larger radii
the anisotropy tends towards \mbox{$\beta\sim0.2-0.3$} for all cusp
strengths, with the less strongly cusped haloes tending to have higher
anisotropy.

Almost all of the $\beta$ profiles share a similar non-monotonic shape. $\beta$ increases
at small $r$, reaching a peak at $r\sim1-1.5 r_{\mathrm h}$, then falls off,
reaching a local minimum at $r\sim2-3 r_{\mathrm h}$, before rising again.
This apparent oscillatory variation of $\beta$ is not associated with any
tidal arms, shells, or any other observed features of the particle 
distribution; it is also not seen to disappear or vary to any great extent
with time. A similar $\beta$ profile can be seen in Figure 11 of 
\cite{BoylanKolchinMa2004}. Further work is required to determine its
cause and importance.

\begin{figure*}
  \centerline{\hfil 
  \resizebox{55mm}{!}{\includegraphics{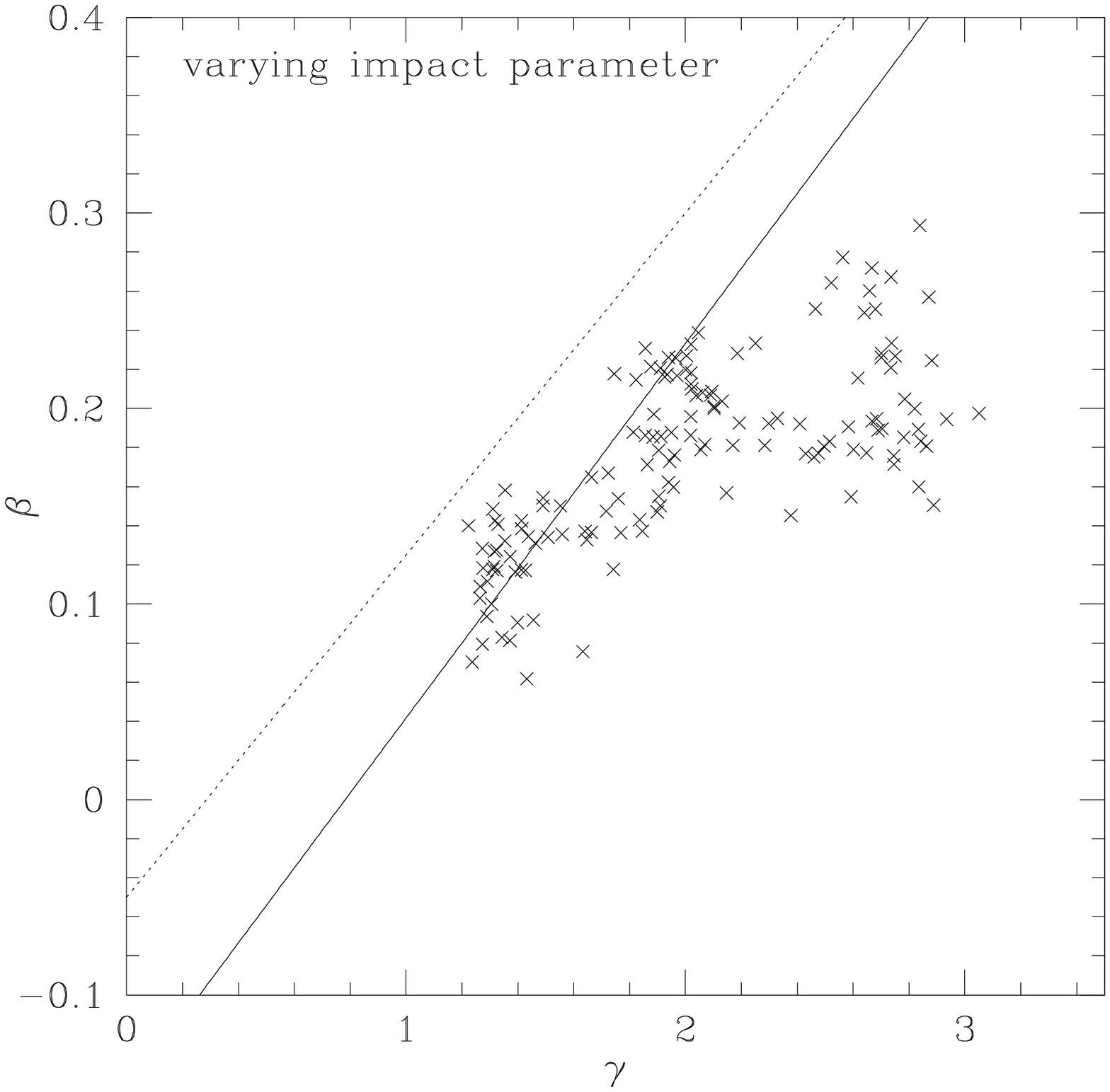}}\hspace{1mm}
  \resizebox{55mm}{!}{\includegraphics{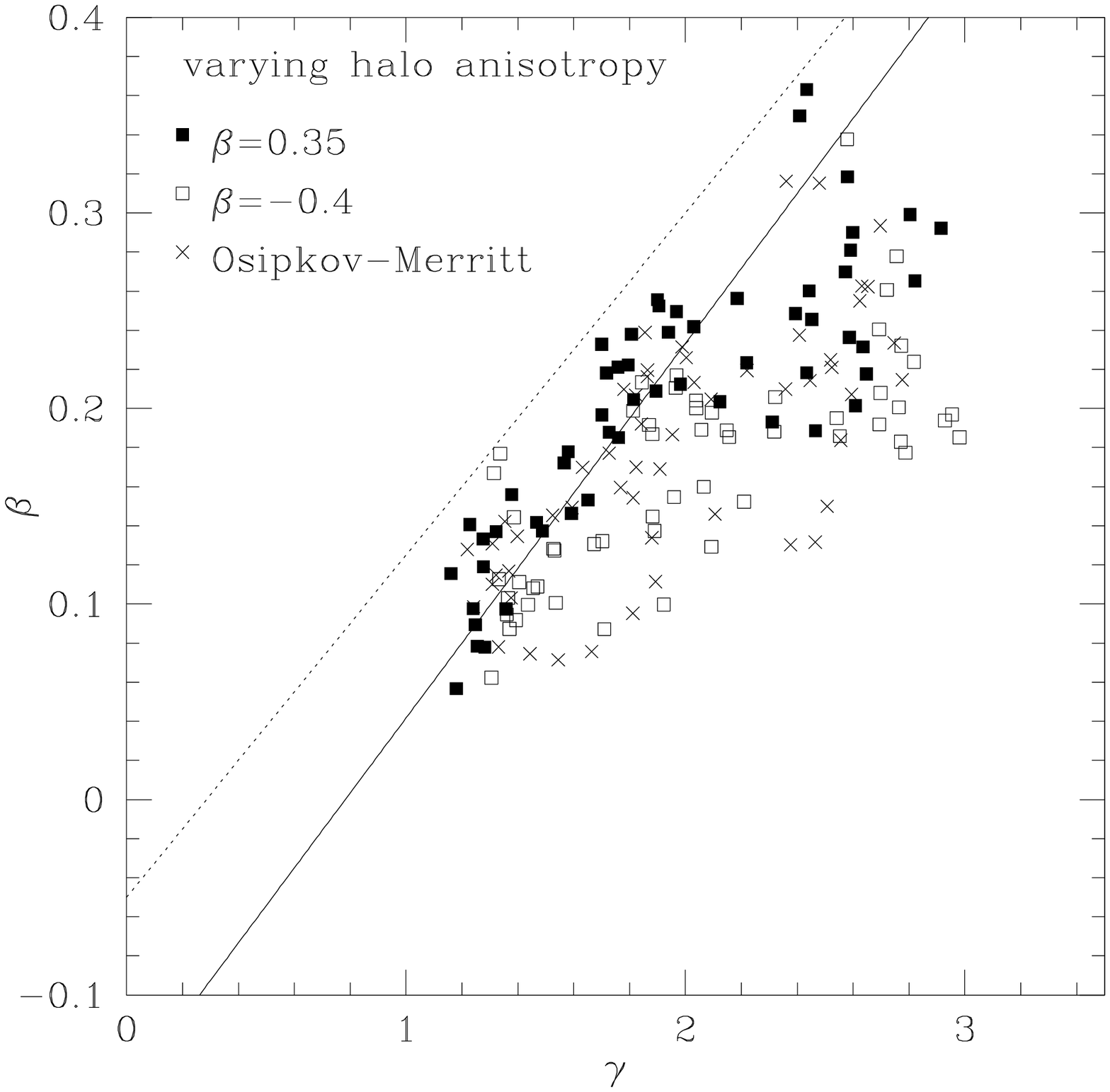}}\hspace{1mm}
  \resizebox{55mm}{!}{\includegraphics{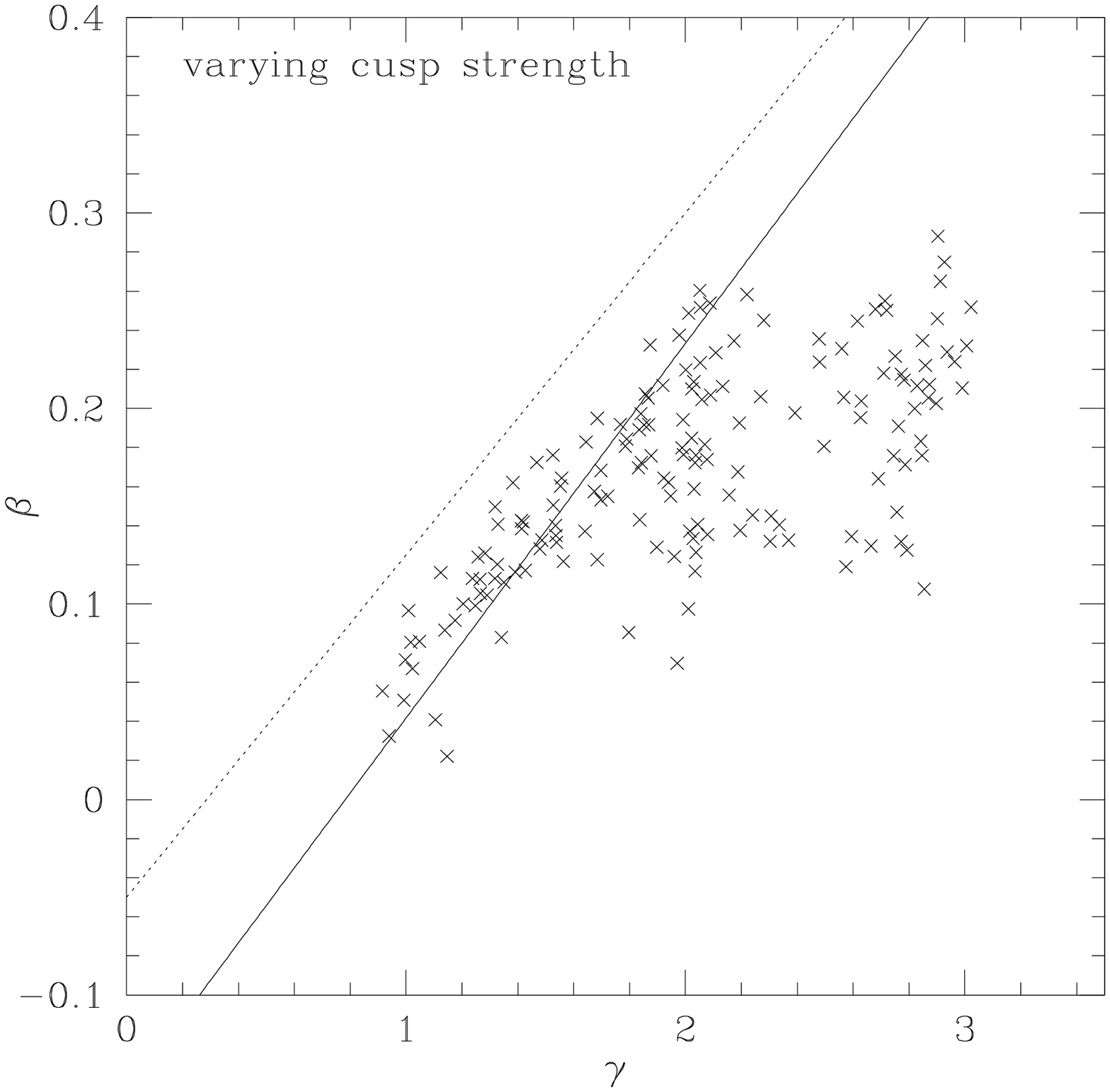}}
}
  \caption{Anisotropy parameter $\beta$ plotted as a function of the local density
        slope $\gamma$ for mergers from suite 2 (\emph{left}), suite 3 (\emph{centre}),
        and suite 4 (\emph{right}). $\beta$ and $\gamma$ are determined for each
        simulation at many points with the radial range $0.1\rh<r<10\rh$.
        The lines drawn in each plot correspond to  
        equation \ref{eq:HM}, with $\xi=1.15$ (\emph{solid line}) and
        $\xi=1.05$ (\emph{dotted}). 
        In the central plot, filled squares correspond to mergers 
        with $\beta=0.35$ initial conditions, open squares are from the $\beta=-0.4$ 
        initial conditions, and crosses for the Osipkov-Merritt halo initial conditions.
        \label{fig:gammabeta}
}       
\end{figure*}

\cite{HansenMoore2006} argued for a universal relationship between local density 
slope $\gamma$ and velocity anisotropy $\beta$, given in Equation~\ref{eq:HM}, 
which  we plot in each 
of the graphs in Figure~\ref{fig:gammabeta}. The data fit the 
suggested relationship for $\gamma\lesssim2$, but do not fit for 
$\gamma\gtrsim2$, where $\beta$ is systematically lower than the 
fitting function. While it is the outer parts of the halo which have $\gamma\gtrsim2$,
and the dynamical time is longer here than in the inner parts of the halo,
this is not simply because the outer halo has yet to reach an equilibrium. 
Simulations which ran for twice as long (until $t=2000\cong28$Gyr scaled 
to the Milky Way) showed identical results. The $\beta=0.35$ 
initial conditions produce remnants which are a close fit to the suggested 
relationship, but the $\beta=-0.4$ initial conditions produce remnants 
which have systematically slightly lower values of $\beta$ for the same $\gamma$.
The fitting formula equation~\ref{eq:HM}, with the smaller value for the 
 free parameter $\xi\simeq1.05$
(plotted as a dotted line in figure~\ref{fig:gammabeta}) does appear to 
provide an upper limit for $\beta$.

We observe that the haloes' net rotation is very small; 
we compare to the velocity dispersion and find that $v_{rot}/\sigma<0.1$ 
in the inner regions in all cases. There is slightly more rotation
at large radii, but still $v_{\mathrm rot}/\sigma<0.2$ in all cases. Bulk rotation
is also near negligible, with the axes of the bulk distribution remaining 
near fixed over a long period of time in all cases.

\subsection{Halo Shape} \label{sec:axes}

\begin{figure}
  \centerline{\resizebox{\hsize}{!}{\includegraphics{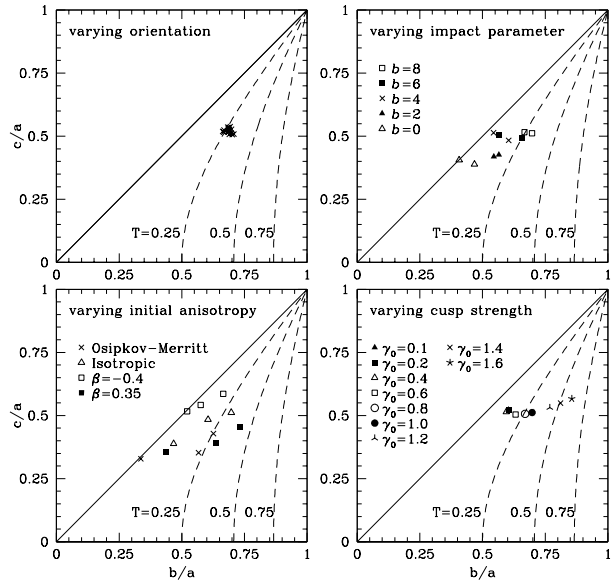}}}
  \caption{Minor ($c/a$) and intermediate ($b/a$) axis ratios for the simulations 
        of all suites of simulations (as labelled).
        In all figures the solid diagonal indicates the position on the plot of a 
        completely prolate halo.
        The dashed lines are of constant triaxiality, as indicated.
        For suites 2 (\emph{top-right}), 3 (\emph{bottom-left})
        and 4 (\emph{bottom-right}) the different symbols represent different
        simulation parameters as labelled. The three points for each initial 
        anisotropy (suite 3) correspond to simulations with impact parameters
        $d=0,3,8$ with, in each case, $b/a_{d=0}<b/a_{d=3}<b/a_{d=8}$.  
        \label{fig:allaxes}
  }
\end{figure}

We calculate the overall shape of the remnant halo through an iterative procedure 
based on diagonalizing the moment of inertia tensor
\begin{equation} \label{eq:MoI}
I_{i,j}=\sum_{\alpha}m_{\alpha}r_{i,\alpha}r_{j,\alpha},
\end{equation}
where the sum is over all particles considered, and $r_{i,\alpha}$ is the $i$ coordinate 
of the $\alpha$th particle. 
We start with a spherical window, centred at the densest point of the halo, 
which contains 50\% of the mass of the halo, 
then determine and diagonalize its moment of inertia tensor. This 
determines the principle axes and the axis ratios of an ellipsoidal window 
which is scaled to contain 
50\% of the halo mass. This is then repeated to find new ellipsoidal windows until the 
results converge such that the volume of the window is conserved to within a fractional 
difference of $10^{-3}$ between iterations. From the ellipsoidal window we determine the
axis lengths $a, b$ \& $c$ with $a\geq b\geq c$, and thus the intermediate ($b/a$) and 
minor ($c/a$) axis ratios. We can then define an ellipsoidal ``radius''
\begin{equation}
\zeta=\sqrt{x^{\prime2}+\frac{a^2y^{\prime2}}{b^2}+\frac{a^2z^{\prime2}}{c^2}}.
\end{equation}
The question of whether an ellipsoidal body is prolate, oblate or triaxial can be 
quantified through the parameter 
\begin{equation}
T=(a^2-b^2)/(a^2-c^2).
\end{equation}
We use this method to enable a like-for-like comparison with the
simulations of \cite{Novaketal2006}, though starting with a spherical window
can cause bias.
\subsubsection{Trends in halo shapes}

The graphs of Figure~\ref{fig:allaxes}, which show the ratios $c/a$ and
$b/a$ (determined as described above) 
for simulations from all of our suites, tell us the following things: 

\begin{enumerate}
\item Remnant haloes are significantly non-spherical ($c/a\lesssim0.55$ 
in all cases shown). They are generally prolate, or at least tending 
towards prolate, in our simulations.
\item The triaxiality of the halo at this scale ($\zeta\sim10\rh$ 
for outermost particles) is barely affected by the initial inclination 
of the disc. 
\item More radial mergers (i.e.~mergers with lower values of 
$d$) tend to have more prolate remnant haloes. This is true whether or
not the initial haloes are isotropic.
\item Tangential anisotropy 
in the initial halo tends to produce haloes that are more prolate, and 
have a higher value of $c/a$ than in the isotropic case. Conversely, 
radial anisotropy  in the initial halo tends to produce haloes 
that are more triaxial, and have a lower value of $c/a$ than in the 
isotropic case. 
\item In the range $0.4<\gamma_0<1.6$, an increasing cusp
strength tends to produce remnants with a greater value of $b/a$. This is not simply due to 
the decreasing size of the window containing 50\% of the mass in haloes
with increasing cusp strength, and is also seen if the volume of the 
window is defined such that it is the same for all cusp strengths.
\end{enumerate}

Like \cite{Novaketal2006} we find that the minor axis of the stellar 
remnant and the major axis of the remnant halo are nearly always 
close to perpendicular. The major axis of the halo is always close to the plane of galaxies' 
original mutual orbit, and the minor axis of the stellar component is nearly
always close to being perpendicular to that plane.


\subsubsection{Ellipticity profiles}\label{sec:axes:prof}
\begin{figure}
  \centerline{\resizebox{\hsize}{!}{\includegraphics{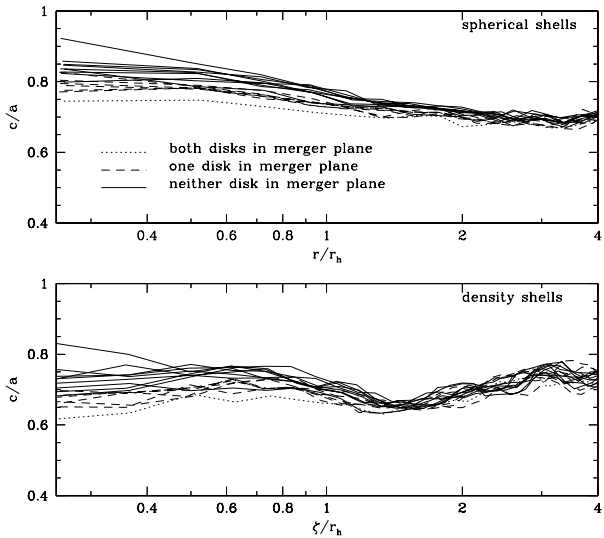}}}
  \caption{Minor axis ratios as a function of radius $r$ (\emph{upper})
        or ellipsoidal radius $\zeta$ (\emph{lower})
        for the simulations of suite 1. The \emph{dotted} 
        lines corresponds to the simulation in which galaxy discs are both oriented 
        in the merger plane, the \emph{dashed} lines correspond to the simulations 
        in which one of the discs was oriented in the merger plane.
        \label{fig:axespro}
  }
\end{figure}

We wish to examine the effect of the orientation of the disc components upon the 
ellipticity of the remnant halo as a function of radius. In an effort
to do this, we find the density at each particle 
by a ``nearest neighbours'' analysis \citep{CasertanoHut1985}, 
divide the particles into shells by density,
and determine the axis ratios and median ellipsoidal radius of each shell. 
The axis ratio of the particles in a shell is found from their moment of 
inertia tensor (Equation~\ref{eq:MoI}). For comparison to \cite{Kazanzidisetal2004}
we also divide the merger remnant halo into spherical shells, and find the axis ratio 
of the particles in those shells. We defined the radius of the spherical shells as 
being the median radius of the particles in it. The two approaches
produced qualitatively similar results, though, as \cite{Athanassoula2006} shows,
using a spherical window introduces a bias towards larger values for $c/a$ and
$b/a$. In Figure~\ref{fig:axespro} we plot the results from the both analyses.

Considered in terms of $c/a$ at the innermost point determined, the flattest 
merger remnant comes from the co-planar merger, and the next 
four flattest remnants are from mergers with one of the discs in the 
orbital plane. The difference in minor axial ratio is relatively small. 
Using the density shells values, at the innermost point 
for the co-planar merger remnant $c/a$ is $\sim0.62$, for mergers with one 
galaxy disc in the orbital plane it's $\sim0.67$, and for mergers with neither 
disc in the orbital plane it's $\sim0.72$. The simulation with orientation 
16 (see Table~1) produces a notably more spherical remnant 
than all the other simulations ($c/a \sim0.83$). Beyond $r\sim1.5\rh$ the difference 
in $c/a$ is negligible. There is no trend apparent in the intermediate axis ratio, $b/a$.
These trends are exactly as one would expect.

It is reasonable to ask whether the fact that these haloes are so far
from spherical invalidates the approach in Section~\ref{sec:dens}, of
using a spherically averaged density profile. In an effort to 
investigate this, we created density profiles based upon the density
shells, found through the method described above. 
These profiles, while suffering somewhat from increased noise,
showed the same behaviour as 
that described in Section~\ref{sec:dens}. This indicates that the use of 
spherically averaged profiles is sufficient and appropriate for the analysis 
of the haloes of merger remnants. 

\section{Discussion}\label{sec:diss}
We have performed a number of suites of simulations, covering a number of
different merger parameters. Across all these simulations we have seen 
that the halo cusp strength is extremely robust against changes caused 
by major mergers, even when there is a centrally concentrated
stellar component. This is in keeping with previous results from halo
only simulations \citep[e.g.][]{KazantzidisZentnerKravtsov2006}; simulations 
with disc components, but initialised using a Maxwellian approximation
\citep[e.g.][]{AcevesVelazquez2006} and analytical arguments 
\citep{Dehnen2005}. 
Thus mergers, at least of gas-less equal mass galaxies, can not solve
the persistent discrepancy between observations of
nearby galaxies, which imply that galactic dark matter haloes have a
density profile with a flat core, and the cosmological standard model,
which predicts that haloes should have a cusp.

In contrast to the density profile, the velocity anisotropy of the halo has very little 
``memory'' of the initial conditions of the merger. In all cases examined,
the velocity anisotropy of the remnant followed the relationship (Equation~\ref{eq:HM})
suggested by \cite{HansenMoore2006} relatively closely for 
$\gamma\lesssim2$, but $\beta$ was systematically lower than predicted
for $\gamma\gtrsim2$. That these results are nearly independent of initial
anisotropy is very encouraging. The overwhelming majority of merger simulations 
to date have been performed with isotropic haloes, while the haloes found in 
cosmological simulations are typically radially anisotropic. A strong dependence
on anisotropy would suggest that the approximation of isotropy in the 
halo was a poor one, which would have cast doubt upon results from simulations
using it. The velocity anisotropy of the halo is clearly a less robust property than
its density profile.

Given the shape of the dark matter haloes observed by, for example 
\cite{Buoteetal2002}, the results of Section~\ref{sec:axes} are of 
particular importance. Axis ratios $c/a$ in the range $0.37\pm0.04$ (as seen
for NGC720) are
seen for all initial halo anisotropies except $\beta=-0.4$, in cases with 
sufficiently small impact parameters. Simulations which incorporate the
effects of gas physics \citep[e.g.][]{Novaketal2006} find remnant haloes 
which are significantly more spherical. It is likely, therefore, that 
galaxies with such highly flattened haloes
were formed from near-radial trajectory, very gas-poor mergers in which collisionless 
dynamics dominate.

That the remnant halo is closest to completely prolate for mergers with low 
impact parameters is unsurprising, and was recognised
in the literature as far back as 1983 \citeauthor{Villumsen1983} in the case of completely 
spherical initial models. In the $d=0$ case, the major axis of the remnant is 
in the same direction as the initial motion of the galaxy centres. 

The dependence of remnant shape 
on initial anisotropy of the haloes is far less trivially understandable, 
but is likely to be related to the radial-orbit instability 
\citep[e.g.][]{Barnes1985,DejongheMerritt1988}. Spherical 
systems with large numbers of radial orbits are known to be 
unstable to deformation towards a barred or triaxial shape. The
radial-orbit instability is also known to reduce the central concentration of models
\citep[e.g.][]{MerrittAguilar1985}, which could explain the slight decrease in 
cusp strength seen in the $\beta=0.35$ model remnants (Figure~\ref{fig:anipro}).

It is known that a steep cusp in a triaxial model causes orbit-scattering 
which can have a significant effect on its shape \citep{ValluriMerritt1998},
reducing the triaxiality. The effect of increasing the cusp strength of our 
haloes is to make the remnants more spherical (less prolate, i.e.~increasing $b/a$), 
but this means that the most 
strongly cusped haloes are \emph{more} triaxial than the least strongly
cusped ones.

Increasing the cusp strength of the halo puts more of the mass of the halo 
at radii smaller than the impact parameter of the merger. This is similar
to the effect of increasing the impact parameter of the merger, which we 
have shown increases $b/a$ for the remnant. It is likely that this 
explains the trend in remnant shape with cusp strength.

\section*{acknowledgements}
PJM acknowledges the support of an EU Marie Curie Fellowship,
and of the UK Particle Physics and Astronomy 
Research Council (PPARC) through a research student fellowship. 
Astrophysics research at the University of Leicester is also supported 
through a PPARC rolling grant.

\bibliographystyle{mn2e}

\bibliography{Mergers}
 
\end{document}